\documentstyle[preprint,aps]{revtex}
\input epsf.tex
\def\DESepsf(#1 width #2){\epsfxsize=#2 \epsfbox{#1}}

\newlength{\dinwidth}
\newlength{\dinmargin}
\begin{document}
\draft
\title {Determination of the CP Violating Weak Phase $\gamma$ \\
        from the Decays $B \to D \pi$}

\author{C. S. Kim\footnote{cskim@mail.yonsei.ac.kr,~~~
http://phya.yonsei.ac.kr/\~{}cskim/} and Sechul
Oh\footnote{scoh@phya.yonsei.ac.kr} }

\address{Department of Physics and IPAP, Yonsei University, Seoul,
120-749, Korea }

\maketitle
\begin{abstract}
\noindent
 We propose a new method to extract the CP violating weak
phase $\gamma$ (the phase of $V_{ub}^*$) in the CKM paradigm of
the Standard Model, using $B^- \rightarrow D^0 \pi^- \rightarrow f
\pi^-$ and $B^- \rightarrow \bar D^0 \pi^- \rightarrow f \pi^-$
decays, where $f$ are final states such as $K^+ \pi^-$, $K^+
\rho^-$,  $K \pi\pi$, {\it etc.} We also study the experimental
feasibility of our new method. With possibility of new phases in
the CKM matrix, we re-examine some of the previously proposed
methods to determine $\gamma$, and find that it would be in
principle possible to identify $\gamma$ and a new phase angle
$\theta$ separately.
\end{abstract}

\newpage
The source for CP violation in the Standard Model (SM) with
\emph{three} generations is a phase in the
Cabibbo-Kobayashi-Maskawa (CKM) matrix \cite{1}.  One of the goals
of $B$ factories is to test the SM through measurements of the
unitarity triangle of the CKM matrix. An important way of
verifying the CKM paradigm is to measure the three angles
\cite{2},
\begin{eqnarray}
\alpha &\equiv& \rm{Arg}[-(V_{td} V^*_{tb})/(V_{ud} V^*_{ub})], \nonumber\\
\beta &\equiv& \rm{Arg}[-(V_{cd}
V^*_{cb})/(V_{td} V^*_{tb})], \nonumber\\
{\rm and}~~~~ \gamma &\equiv& \rm{Arg}[-(V_{ud} V^*_{ub})/(V_{cd}
V^*_{cb})],
\end{eqnarray}
of the unitarity triangle independently of many experimental
observables and to check whether the sum of these three angles is
equal\footnote{ The sum of those three angles, defined as the
intersections of three lines, would be always equal to 180$^0$,
even though the three lines may not be closed to make a triangle.}
to 180$^0$, as it should be in the paradigm.  The angle $\beta$
can be determined unambiguously by measuring the time-dependent CP
asymmetry in the gold-plated mode $B \rightarrow J/\psi K_S$
\cite{3}.  The angle $\alpha$ can be extracted to a reasonable
accuracy through the study of CP asymmetry in $B \rightarrow \pi
\pi$, combined with the isospin analysis to remove the involved
penguin contamination \cite{4}, and through decays of $B \to \rho
\pi$ \cite{5} and $B \to a_0 \pi$ \cite{6}.  It is well known that
among the three angles, $\gamma$ would be the most difficult to
determine in experiment.  There have been a lot of works to
propose methods measuring $\gamma$ using $B$ decays, but at
present there is no gold-plated way to determine this angle. In
particular, a class of methods using $B \rightarrow D K$ decays
have been proposed \cite{7,8,9,10,11,11a}. All the methods
presented in Refs. \cite{7,8,9,10,11,11a} have assumed the
\emph{unitarity} of the CKM matrix and,  therefore, only one
independent phase angle in the CKM matrix.

We present a new method for determining $\gamma$, which is similar
to the Atwood-Dunietz-Soni (ADS) method \cite{9}, but we use $B
\rightarrow D \pi$ decays instead of $B \rightarrow D K$ decays
used in the ADS method.  The CLEO Collaboration have observed
\cite{17} that the branching ratio for $B^- \rightarrow D^0 \pi^-$
is much larger than that for $B^- \rightarrow D^0 K^-$,
\begin{eqnarray}
{{\cal B}(B^- \rightarrow D^0 K^-) \over {\cal B}(B^- \rightarrow
D^0 \pi^-)} = \rm{0.055} \pm \rm{0.014} \pm \rm{0.005} \; .
\label{cleo}
\end{eqnarray}
We consider the decay processes $B^- \rightarrow D^0 \pi^-
\rightarrow f \pi^-$, $B^- \rightarrow \bar D^0 \pi^- \rightarrow
f \pi^-$ and their CP-conjugate processes, where $D^0$ and $\bar
D^0$ decay into common final states $f = K^+ \pi^-$, $K^+ \rho^-$,
$K \pi\pi$, and so forth. We note that the decay mode $B^-
\rightarrow \bar D^0 \pi^-$ is severely suppressed relative to the
mode $B^- \rightarrow D^0 \pi^-$, and this fact causes serious
experimental difficulties in using $B^- \rightarrow \bar D^0
\pi^-$ decay for the Gronau-London-Wyler (GLW) method \cite{7}.
However, in our method one needs not to perform the difficult task
of measuring the branching ratio for $B^- \rightarrow \bar D^0
\pi^-$, similar to the case of the ADS method. The detailed
experimental feasibility for our new method and the ADS method is
given later.

Note that the decay amplitudes of $B^- \rightarrow D^0 \pi^-$ and
$D^0 \rightarrow f$ contain the CKM factors $V^*_{ud} V_{cb}$ and
$V^*_{cd} V_{us}$, respectively, while the amplitudes of $B^-
\rightarrow \bar D^0 \pi^-$ and $\bar D^0 \rightarrow f$ contain
the CKM factors $V^*_{cd} V_{ub} = |V^*_{cd} V_{ub}| e^{-i
\gamma}$ and $V^*_{ud} V_{cs} = |V^*_{ud} V_{cs}|$, respectively.
We define the following quantities: $(i = 1,2)$
\begin{eqnarray}
a &=& A(B^-\rightarrow D^0 \pi^-) =
   |A(B^- \rightarrow D^0 \pi^-)| e^{i \delta_a},   \nonumber \\
b &=& A(B^- \rightarrow \bar D^0 \pi^-) = |A(B^- \rightarrow \bar
D^0 \pi^-)|
   e^{-i \gamma} e^{i \delta_b}, \nonumber \\
c_i &=& A(D^0 \rightarrow f_i) = |A(D^0 \rightarrow f_i)|
   e^{i \delta_{c_i}}, \nonumber \\
c_i^{\prime} &=& A(D^0 \rightarrow \bar f_i) = |A(D^0 \rightarrow
\bar f_i)|
    e^{i \delta_{c^\prime_i}}, \nonumber \\
d_i &=& A(B^- \rightarrow [f_i] \pi^-), \label{abcd}
\end{eqnarray}
where $A$ denotes the relevant decay amplitude and $\delta$'s are
the relevant strong rescattering phases.  Here $[f_i]$ in $d_i$
denotes that $f_i$ originates from a $D^0$ or $\bar D^0$ decay.
Similarly, we also define $\bar a$, $\bar b$, $\bar c_i$, $\bar
c_i^{\prime}$ and $\bar d_i$ as the CP-conjugate decay amplitudes
corresponding to $a$, $b$, $c_i$, $c_i^{\prime}$ and $d_i$,
respectively, such as $\bar d_i = A(B^+ \rightarrow [\bar f_i]
\pi^+)$, {\it etc.} Note that $|x| = |\bar x|$ with $x =
a,b,c_i,c_i^{\prime}$, but in general $|d_i| \neq |\bar d_i|$, as
shown below. Then, the amplitude $d_i$ can be written as
\begin{eqnarray}
d_i &=& A(B^- \rightarrow D^0 \pi^-) A(D^0 \rightarrow f_i)
+A(B^- \rightarrow \bar D^0 \pi^-) A(\bar D^0 \rightarrow f_i) \nonumber\\
&=& a c_i + b \bar c_i^{\prime}  \nonumber \\ &=& |a c_i| e^{i
(\delta_a +\delta_{c_i})} +|b \bar c_i^{\prime}| e^{-i \gamma}
e^{i (\delta_b +\delta_{c_i^{\prime}})}. \label{di}
\end{eqnarray}
Thus, $|d_i|^2$ and $|\bar d_i|^2$ are given by
\begin{eqnarray}
|d_i|^2 = |a c_i|^2 +|b \bar c_i^{\prime}|^2 +2 |a b c_i \bar
c_i^{\prime}| \cos(\gamma +\Delta_i), \nonumber \\ |\bar d_i|^2 =
|a c_i|^2 +|b \bar c_i^{\prime}|^2 +2 |a b c_i \bar c_i^{\prime}|
\cos(\gamma -\Delta_i), \label{dd}
\end{eqnarray}
where $\Delta_i = \delta_a -\delta_b +\delta_{c_i}
-\delta_{c_i^{\prime}}$. We see that $|d_i| \neq |\bar d_i|$,
unless $\Delta_i = n \pi$ ($n =0, 1, ...$). The expressions in Eq.
(\ref{dd}) represent four equations for $i =1,2$. Now let us
assume that the quantities $|a|$, $|c_i|$, $|c_i^{\prime}|$,
$|d_i|$ and $|\bar d_i|$ are measured by experiment, but $|b|$ is
unknown. Then there are the four unknowns $|b|$, $\gamma$,
$\Delta_1$, $\Delta_2$ in the above four equations.  By solving
the equations one can determine $\gamma$, as well as the other
unknowns such as $|b|=|A(B^- \rightarrow \bar D^0 \pi^-)|$.

In the ADS method \cite{9}, $a$, $b$ and $d_i$ in Eq. (\ref{abcd})
are replaced by
\begin{eqnarray}
a &=& A(B^- \rightarrow D^0 K^-) = |A(B^- \rightarrow D^0 K^-)|
   e^{i \delta_a},   \nonumber \\
b &=& A(B^- \rightarrow \bar D^0 K^-) = |A(B^- \rightarrow \bar
D^0 K^-)|
   e^{-i \gamma}  e^{i \delta_b},  \nonumber \\
d_i &=& A(B^- \rightarrow [f_i] K^-). \label{ads}
\end{eqnarray}
Then, $|d_i|^2$ and $|\bar d_i|^2$ can be expressed by the same
form as in Eq. (\ref{dd}).  Therefore, the phase $\gamma$ can be
determined by solving the four equations (for $i=1,2$) with four
unknowns $|b|$, $\gamma$, $\Delta_1$, $\Delta_2$. The impact on
the ADS method due to the large $D^0 - \bar D^0$ mixing from new
physics has been studied in Ref. \cite{18}.

Now we study the experimental feasibility of our new method and
the ADS method, by solving Eq. (\ref{dd}) analytically,
\begin{eqnarray}
\cos(\gamma + \Delta_i)&=&{{|d_i|^2 - |a c_i|^2 - |b \bar
c_i^\prime|^2} \over
 {2 |a c_i b \bar c_i^\prime|}} , \nonumber \\
\cos(\gamma - \Delta_i)&=&{{|\bar d_i|^2 - |a c_i|^2 - |b \bar
c_i^\prime|^2} \over
 {2 |a c_i b \bar c_i^\prime|}} .
\label{cosine}
\end{eqnarray}
To make a rough numerical estimate of the possible statistical
error on determination of $\gamma$, we use the experimental
result, Eq. (\ref{cleo}), and the mean values for the CKM
elements;
\begin{eqnarray}
{\cal B}(B^- &\to& D^0 \pi^-) : {\cal B}(B^- \to D^0 K^-)
: {\cal B}(B^- \to \bar D^0 K^-) : {\cal B}(B^- \to \bar D^0 \pi^-) \nonumber \\
 &\simeq& |V_{cb} V^*_{ud}|^2 : |V_{cb} V^*_{us}|^2
 : |V_{ub} V^*_{cs}/N_c|^2 :
           |V_{ub} V^*_{cd}/N_c|^2 \nonumber \\
 &\approx& A^2 \lambda^4 \times (1 ~:~ \lambda^2 ~:~ \lambda^2/36 ~:~
\lambda^4/36) \nonumber \\  &\approx& 100 ~:~ 5 ~:~ 0.15 ~:~ 0.007
\nonumber
\\ &\sim& {\cal O}(100) ~:~ {\cal O}(10) ~:~ {\cal O}(0.1) ~:~
{\cal O}(0.01) ,  \label{bbbb}
\end{eqnarray}
where we used $|V_{ub}/V_{cb}| \approx \lambda/2$, the
color-suppression factor $N_c=3$, $\lambda=\sin\theta_C=0.22$, and
$A=V_{cb}/\lambda^2$ is a Wolfenstein parameter. In order to
consider the decay parts, $c_i,~\bar c^\prime_i$, we choose the
modes such as $|\bar c^\prime_i| >> |c_i|$, $e.g.$,
\begin{eqnarray}
&{}& |c(D^0 \to K^+ \pi^-)|^2 ~:~ |\bar c^\prime (\bar D^0 \to K^+
\pi^-)|^2 \nonumber \\
 &=& {\cal B}(D^0 \to K^+ \pi^-) : {\cal B}(\bar D^0 \to K^+
 \pi^-) \nonumber \\
 &=& (1.5 \pm 0.3) \times 10^{-4} : (3.8 \pm 0.1) \times 10^{-2}
 \nonumber \\
 &\sim& ~{\cal O}(1) ~:~ {\cal O}(100),
 \label{ccprime}
\end{eqnarray}
which makes
\begin{eqnarray}
&{}& |a c_i|^2(\pi) ~:~ |a c_i|^2(K) ~:~ |b \bar c_i^\prime|^2(K)
~:~
|b \bar c_i^\prime|^2(\pi) \nonumber \\
 &\propto& ~{\cal O}(100) ~:~ {\cal O}(10) ~:~ {\cal O}(10) ~:~ {\cal
 O}(1).
 \label{acacbcbc}
\end{eqnarray}
Therefore, if we assume the  1 \% level precision  in the
experimental determination for product of branching  ratios
\footnote{At present, the experimental data for this product of
branching ratio is given in about 29 \% level precision, using a
few times $10^6$ $B$ mesons \cite{prd}. Thus, to obtain the data
in 1 \% level precision, one needs to have about $10^9$ $B$'s,
which can be achieved at hadronic $B$ experiments such as BTeV and
LHC-B, where more than $10^{10}$ $B$ mesons will be produced per
year. }, $e.g.$, $\Delta[{\cal B}(B^- \to D^0 \pi^-) \times {\cal
B}(D^0 \to K^+ \pi^-)] = 1 \%$, then we can set the numerical
values, for $B^\pm \to D(\to f_i)\pi^\pm$,
\begin{eqnarray}
|a c_i|^2(\pi)\approx 100 \pm 1, &{}&~ |b \bar
c_i^\prime|^2(\pi)\approx 1\pm 0.1 ~, \label{acbcpi}
\end{eqnarray}
and for $B^\pm \to D(\to f_i) K^\pm$,
\begin{eqnarray}
|a c_i|^2(K)\approx 10 \pm 0.3,  &{}&~ |b \bar
c_i^\prime|^2(K)\approx 10 \pm 0.3~. \label{acbck}
\end{eqnarray}
Then,  we can make rough estimate for the statistical error from
Eq. (\ref{cosine}) as
\begin{eqnarray}
\Delta[\cos(\gamma +\theta \pm \Delta_i)(B^\pm \to D(\to
f_i)\pi^\pm)] &\sim& 0.1 ~,
\nonumber \\
\Delta[\cos(\gamma \pm \Delta_i)(B^\pm \to D(\to f_i)K^\pm)]
&\sim& 0.05 ~. \label{delcosine}
\end{eqnarray}
We find  the ADS method can give approximately twice better
precision statistically for determination of $\gamma$, compared to
our new method.

In fact, there is a general theorem \cite{london}:
\begin{eqnarray}
N_B~  \propto  1 / ( {\cal B}(B \to f) A_f^2 )~,
\end{eqnarray}
where $N_B$ is the number of $B$ mesons needed, ${\cal B}$ the
branching ratio of a decay mode, $B \to f$, and $A_f$ the relevant
asymmetry. Now, as shown in Eq. (\ref{cleo}), ${\cal B}$(ADS
method)/${\cal B}$(our method) $\simeq 0.05$.  To determine the
relevant asymmetry $A_f$, one has to calculate the following:
\begin{eqnarray}
A_f &=& \frac{ |d_i|^2 - |\bar d_i|^2}{|d_i|^2 + |\bar d_i|^2}
\nonumber \\
 &=& \frac{-2 |abc_i \bar c_i^{\prime}| \sin\gamma
\sin\Delta_i}{|ac_i|^2 +|b \bar c_i^{\prime}|^2 +2 |abc_i \bar
c_i^{\prime}| \cos\gamma \cos\Delta_i} \nonumber \\
&\sim& \frac{2 |abc_i \bar c_i^{\prime}|}{|ac_i|^2 +|b \bar
c_i^{\prime}|^2}.
\end{eqnarray}
For simplicity, we have considered the maximum asymmetry in both
methods.  Using the experimental values given in Eqs.
(\ref{cleo},~\ref{bbbb} $-$ \ref{acbck}), we can easily get
$A_f$(ADS method)/$A_f$(our method) $\simeq (5 - 10)$.  Therefore,
$N_B$(ADS method)/$N_B$(our method) $\sim (1 - 0.2)$, which is
exactly consistent with the above prediction, Eq.
(\ref{delcosine}), where we have predicted the possible precision
with the same number of $B$ mesons.

We note that our new method may have other advantages:
\begin{itemize}
\item
The values of $|d_i|^2~{\rm and}~|\bar d_i|^2 \propto {\cal
B}(B^\pm \to [f_i]\pi^\pm)$ are an order of magnitude bigger than
$|d_i|^2~{\rm and}~|\bar d_i|^2 \propto {\cal B}(B^\pm \to
[f_i]K^\pm)$. Therefore, if the present asymmetric $B$ factories
of Belle and Babar can produce only a handful of such events
because of the limited detector and trigger efficiencies, our new
method may be the first measurable option.
\item
Systematic errors could be much smaller for our new method due to
the final state particle identification, $i.e.$ fewer number of
the final state pions due to $K \to \pi \pi$, and the
reconstruction of $K$.
\end{itemize}

Now we would like to make comments on new physics effects on
determination of weak phase $\gamma$. There can be two independent
approaches to find out new physics beyond the SM, if it exists.
\begin{itemize}
\item
The unitarity of CKM matrix can be assumed. In this case, new
physics effects can only come out from new virtual particles  or
through new interactions in penguin or box diagrams in $B$ meson
decays. If this is the case, all the methods which we mentioned
above will extract the exactly same $\gamma$.
\item
The CKM matrix can be generalized to the non-unitary matrix. In
this case, new physics effects can appear even in tree diagram
decays. And the values of $\gamma$ extracted from each method can
be different. Therefore, we will describe in more detail for this
second case.
\end{itemize}

In fact, in models beyond the SM, the CKM matrix may not be
unitary; for instance, in a model with an extra down quark singlet
(or more than one), or an extra up quark singlet, or both up and
down quark singlets, the CKM matrix is no longer unitary
\cite{16,16a}. If the unitarity constraint of the CKM matrix is
removed, the generalized CKM matrix possesses 13 independent
parameters (after absorbing 5 phases to quark fields) -- it
consists of 9 real parameters and \emph{4 independent phase
angles}. The generalized CKM matrix can be parameterized as
\cite{12}
\begin{eqnarray}
\left( \matrix{ |V_{ud}| & |V_{us}| & |V_{ub}|e^{i \delta_{13}}
\cr
                |V_{cd}| & |V_{cs}|e^{i \delta_{22}} & |V_{cb}| \cr
|V_{td}|e^{i \delta_{31}} & |V_{ts}| & |V_{tb}|e^{i \delta_{33}}
\cr } \right). \label{ckm}
\end{eqnarray}
With the possibility of the non-unitary CKM matrix, one has to
carefully examine the effects from the non-unitarity on the
previously proposed methods where the unitarity of the CKM matrix
was assumed to test the SM for CP violation.  From now on, we set
$\gamma \equiv -\delta_{13}$ and $\theta \equiv -\delta_{22}$.

In the parameterization given in Eq. (\ref{ckm}), using our
method, $c_i^{\prime}$ in Eq. (\ref{abcd}) is replaced by
\begin{eqnarray}
c_i^{\prime} = |A(D^0 \rightarrow \bar f_i)|
   e^{i \theta}   e^{i \delta_{c^\prime_i}}.
   \label{ci}
\end{eqnarray}
This leads to the result that the phase $\gamma$ in the
expressions for $|d_i|^2$ and $|\bar d_i|^2$ in Eq. (\ref{dd})
should be replaced by $(\gamma +\theta)$.  Therefore, in this
case, our method can measure the non-unitary phase $(\gamma
+\theta)$.

In the ADS method, besides $c_i^{\prime}$ is changed into the one
in Eq. (\ref{ci}), the phase $\gamma$ in $b$ is also replaced by
$(\gamma -\theta)$.  As a result, the new phase $\theta$ is
automatically cancelled to disappear in the expressions for
$|d_i|^2$ and $|\bar d_i|^2$.  Thus, the ADS method would still
measure $\gamma$ that \emph{is the phase of} $V_{ub}^*$.

The GLW method \cite{7} was suggested for extracting $\gamma$ from
measurements of the branching ratios of decays $B^{\pm}
\rightarrow D^0 K^{\pm}$, $B^{\pm} \rightarrow \bar D^0 K^{\pm}$
and $B^{\pm} \rightarrow D_{CP} K^{\pm}$, where $D_{CP}$ is a CP
eigenstate. However, the GLW method suffers from serious
experimental difficulties, mainly because the process $B^-
\rightarrow \bar D^0 K^-$ (and its CP conjugate process $B^+
\rightarrow D^0 K^+$) is difficult to measure in experiment. That
is, the rate for the CKM-- and color--suppressed process $B^-
\rightarrow \bar D^0 K^-$ is suppressed by about two orders of
magnitudes relative to that for the CKM-- and color--allowed
process $B^- \rightarrow D^0 K^-$, and it causes experimental
difficulties in identifying $\bar D^0$ through $\bar D^0
\rightarrow K^+ \pi^-$ since doubly Cabibbo--suppressed $D^0
\rightarrow K^+ \pi^-$ following $B^- \rightarrow D^0 K^-$
strongly interferes with $\bar D^0 \rightarrow K^+ \pi^-$
following the rare process $B^- \rightarrow \bar D^0 K^-$. With
the non-unitary CKM matrix, this method would measure the angle
$(\gamma - \theta)$ (see Figure 1), instead of $\gamma$ as
originally proposed in Ref. \cite{7}.

In Ref. \cite{11} two groups, Gronau and Rosner (GR), Jang and Ko
(JK), proposed a method to extract $\gamma$ by exploiting
Cabibbo--allowed decays $B \rightarrow D K^{(*)}$ and using the
isospin relations.  In the GR/JK method, the decay modes $B
\rightarrow DK$ with the quark process $b \rightarrow u \bar c s$
contain the CKM factor $|V_{ub} V^*_{cs}| e^{-i (\gamma-\theta)}$
and their amplitudes can be written as
\begin{eqnarray}
A(B^- \rightarrow \bar D^0 K^-) &=& \left( {1 \over 2} A_1 e^{i
\delta_1}
+{1 \over 2} A_0 e^{i \delta_0} \right) e^{-i (\gamma -\theta)}, \nonumber \\
A(B^- \rightarrow D^- \bar K^0) &=& \left( {1 \over 2} A_1 e^{i
\delta_1}
-{1 \over 2} A_0 e^{i \delta_0} \right) e^{-i (\gamma -\theta)}, \nonumber \\
A(\bar B^0 \rightarrow \bar D^0 \bar K^0) &=& A_1 e^{i \delta_1}
     e^{-i (\gamma -\theta)},
\label{AAA}
\end{eqnarray}
where $A_i$ and $\delta_i$ denote the amplitude and the strong
re-scattering phase for the isospin $i$ state. Note that the weak
phase angle $(\gamma -\theta)$ appears in Eq. (\ref{AAA}) rather
than $\gamma$ as in Ref. \cite{11}. In this method, three
triangles are drawn to extract $2 \gamma$, using the isospin
relation
\begin{eqnarray}
 A(B^- \rightarrow \bar D^0 K^-) +A(B^- \rightarrow D^- \bar
K^0) = A(\bar B^0 \rightarrow \bar D^0 \bar K^0)
\end{eqnarray}
and the following relations
\begin{eqnarray}
A(B^- \rightarrow D_1 K^-)
 &=& A(\bar B^0 \rightarrow D_1 \bar K^0) +{1 \over \sqrt{2}}
   A(\bar B^0 \rightarrow D^+ K^-),  \nonumber \\
 A(B^+ \rightarrow D_1 K^+)
&=& A(B^0 \rightarrow D_1 K^0) +{1 \over \sqrt{2}} A(B^0
\rightarrow D^- K^+),
\end{eqnarray}
where $D_1$ is a CP eigenstate of $D$ meson, defined by $D_1 = {1
\over \sqrt{2}} (D^0 + \bar D^0)$.  The appearance of $(\gamma
-\theta)$ in Eq. (\ref{AAA}) results in extraction of $2 (\gamma
-\theta)$, by this method, rather than $2 \gamma$ as in the above
reference, as shown in Fig. 1.

In Fig. 1, $A(B^- \rightarrow D_1 K^-)$, $A(B^- \rightarrow \bar
D^0 K^-)$ and a thick solid line form a triangle, and $A(B^+
\rightarrow D_1 K^+)$, $A(B^+ \rightarrow D^0 K^+)$ and the other
thick solid line form another triangle. These two triangles are
those used in the GLW method and the thick solid lines correspond
to ${1 \over \sqrt{2}} A(B^- \rightarrow \bar D^0 K^-)$ and ${1
\over \sqrt{2}} A(B^+ \rightarrow D^0 K^+)$, respectively, which
are very difficult to measure in experiment. Obviously, assuming
that the GLW method works, the method would extract $2 (\gamma
-\theta)$ rather than $2 \gamma$. We note that the impact on the
GLW method from the sizable $D^0 - \bar D^0$ mixing via new
physics effects has been also investigated in Ref. \cite{19}.

In conclusion, we have presented a new method to determine $\gamma
= \rm{Arg}(V^*_{ub})$ in the CKM paradigm of the SM, using $B^-
\rightarrow D^0 \pi^- \rightarrow f \pi^-$ and $B^- \rightarrow
\bar D^0 \pi^- \rightarrow f \pi^-$ decays, where $f = K^+ \pi^-$,
$K^+ \rho^-$, $K \pi\pi$, {\it etc.}  The experimental feasibility
of our method, comparing with the ADS method, has been studied.
Although experimentally challenging at present asymmetric $B$
factories of Belle and Babar, the analysis using our method can be
carried out in details at hadronic $B$ experiments such as BTeV
and LHC-B, where more than $10^{10}$ $B$ mesons will be produced
per year.

With possibility of new phases in the CKM matrix, we have
re-examined some of the previously proposed methods for
determining the weak phase $\gamma$ using $B \rightarrow DK$ or $B
\rightarrow D \pi$ decays.  We have shown that our method would
extract $(\gamma + \theta)$ with the new phase $\theta$.  The ADS
method would measure $\gamma$, while the GLW method or the GR/JK
method would measure $(\gamma -\theta)$.  Thus, if one uses the
above methods independently and compares the results, it would be
in principle possible to identify $\gamma$ and $\theta$
separately. If this is the case and $\theta$ is not negligible,
this would be a clear indication of the new phase in the CKM
matrix, {\it i.e.} an effect from new physics.  We note that, in
fact, inconsistencies between the values of $\gamma$ can arise
from final state interactions (FSIs) which are known to be
important in $K$ decays and, consequently, may modify quark-level
description of $B$ decays. Therefore, there is a possibility that
FSIs may cause a problem in identifying the new phase $\theta$ by
using various methods mentioned above.

\section*{Acknowledgments}
The work of C.S.K. was supported in part by  Grant No.
2000-1-11100-003-1 and SRC Program of the KOSEF, and in part by
the KRF Grants, Project No. 2000-015-DP0077. The work of S.O. was
supported by the KRF Grants, Project No. 2000-015-DP0077 and by
the BK21 Project.

\newpage
\begin{figure}[htb]
\vspace{1 cm}

\centerline{ \DESepsf(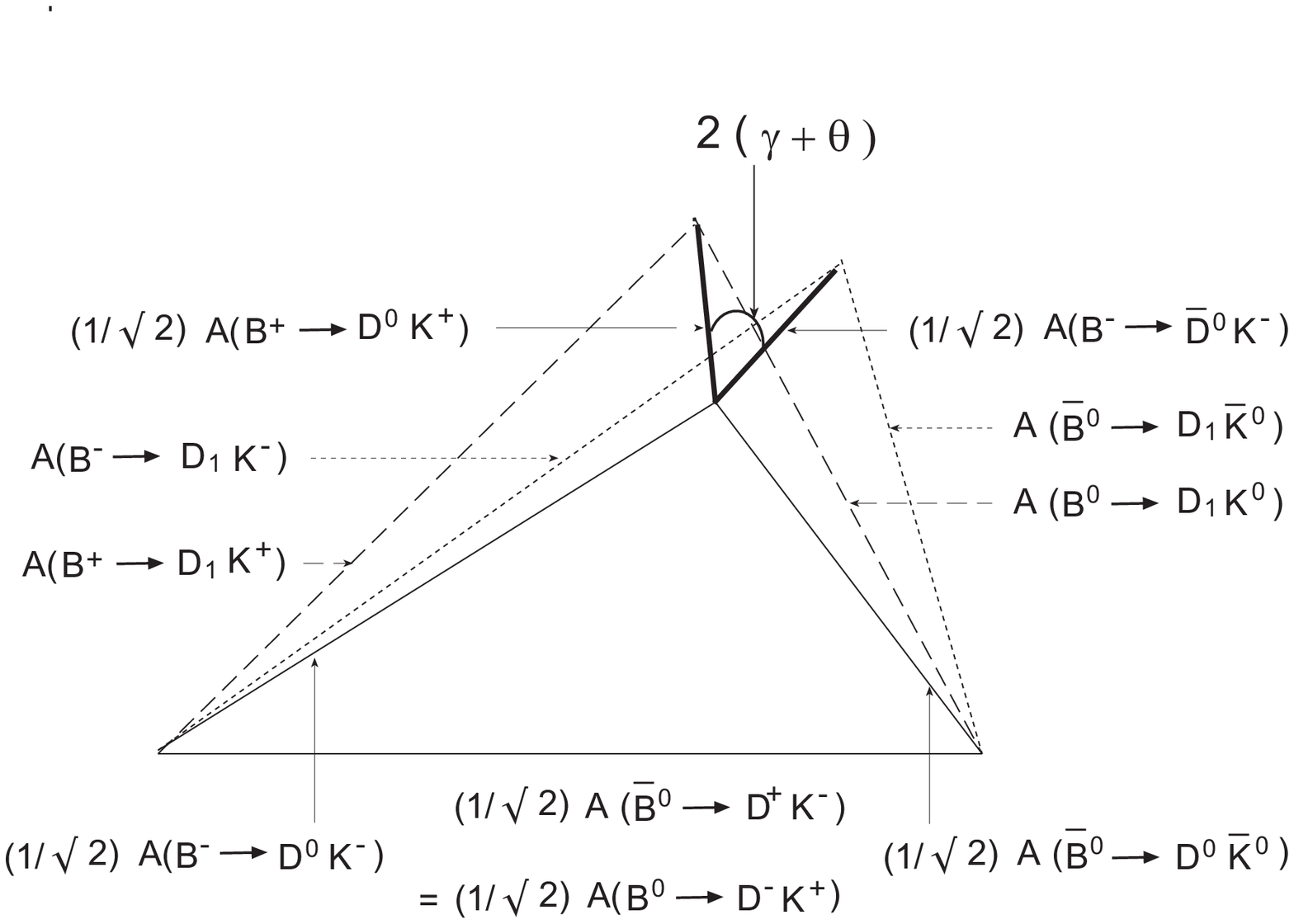 width 12 cm) }
\smallskip
\caption {Three triangles constructed from various $B \rightarrow
DK$ decays [11]. With a new phase in the CKM matrix, one would
extract $(\gamma +\theta)$ rather than $\gamma$ using the method
in Ref. [11], as shown in the figure.}
\end{figure}
\end{document}